# Scholarly use of social media and altmetrics: a review of the literature

Cassidy R. Sugimoto[1], Sam Work[2], Vincent Larivière[2,3] & Stefanie Haustein[2]

[1] sugimoto@indiana.edu
School of Informatics and Computing, Indiana University Bloomington
901 E. 10th St. Bloomington, IN 47408 (USA)

[2] sam.work.15@ucl.ac.uk, stefanie.haustein@umontreal.ca,
École de bibliothéconomie et des sciences de l'information, Université de Montréal
C.P. 6128, Succ. Centre-Ville, Montréal, QC. H3C 3J7 (Canada)

[3] vincent.lariviere@umontreal.ca
Observatoire des sciences et des technologies (OST), Centre interuniversitaire de recherche sur la science et la technologie (CIRST), Université du Québec à Montréal
CP 8888, Succ. Centre-Ville, Montréal, QC. H3C 3P8, (Canada)

## Abstract

Social media has become integrated into the fabric of the scholarly communication system in fundamental ways: principally through scholarly use of social media platforms and the promotion of new indicators on the basis of interactions with these platforms. Research and scholarship in this area has accelerated since the coining and subsequent advocacy for altmetrics—that is, research indicators based on social media activity. This review provides an extensive account of the state-of-the art in both scholarly use of social media and altmetrics. The review consists of two main parts: the first examines the use of social media in academia, examining the various functions these platforms have in the scholarly communication process and the factors that affect this use. The second part reviews empirical studies of altmetrics, discussing the various interpretations of altmetrics, data collection and methodological limitations, and differences according to platform. The review ends with a critical discussion of the implications of this transformation in the scholarly communication system.

## 1 Introduction

By all accounts, scholarly communication appears to be subject to perpetual revolution over the past decades. Harnad (1991) heralded electronic publishing as a fourth major *cognitive revolution* on par with the advent of language, writing, and print. Cronin (2012), taking a less radical view, mused that the proliferation of new forms of scholarly communication was evidence of a *velvet revolution*. Nielsen (2012) describes the *open science revolution* that is happening in are era of networked science. Bornmann (2016) argues that the increasing emphasis on measuring societal impact—and the opportunity to do so using electronic means—provides a *taxonomic revolution* for bibliometrics. The theme that permeates all these conceptions of revolution is a shift towards greater visibility and heterogeneity. In recent years, the discourse has focused on the role of social media[1] in increasing the visibility of scholars and scholarship and offering new vehicles for dissemination (e.g., Van Noorden (2014)). Concurrently, the demand—by research funders and administrators—for indicators of scientific and technological activity has never been so high, especially with regard to demonstrating the value of research to a broader audience (Higher Education Funding Council for England, 2011; Nicholas et al., 2015; Piwowar, 2013b; Viney, 2013; Wilsdon et al., 2015). In this context, social media platform have quickly been identified as a potential source to measure the impact—on both science and society—of scholarly research (Priem, Taraborelli, Groth, & Neylon, 2010), which has led to the creation of a new family of science indicators, labeled *altmetrics*.

One of the central issues associated with altmetrics (short for alternative metrics) is the identification of communities engaging with scholarly content on social media (Haustein, Bowman, & Costas, 2015; Neylon, 2014; Tsou, Bowman, Ghazinejad, & Sugimoto, 2015). It is thus of central importance to understand the uses and users of social media in the context of scholarly communication. Although there have been a few targeted reviews (e.g., Holmberg, 2015; Weller, 2015), a comprehensive literature review which brings together research on scholarly social media use and its metrics is still lacking. This review addresses this gap by providing an overview of the literature on the use of social media in academia and in scholarly communication and on the metrics derived from such activity.

The variation in categorizations of social media and the non-exclusivity of platforms within these categorization schemes makes comparisons of studies on social media use problematic. Even in the academic context, definitions and classifications differ; however, most identify the following major categories: social networking, social bookmarking, blogging, microblogging, wikis, and media and data sharing (Gu & Widén-Wulff, 2011; Rowlands, Nicholas, Russell, Canty, & Watkinson, 2011; Tenopir et al., 2013). Some also consider conferencing, collaborative authoring, scheduling and meeting tools (Rowlands et al., 2011) or RSS and online documents (Gu & Widén-Wulff, 2011; Tenopir et al., 2013) as social media. The landscape of social media, as well as that of altmetrics, is constantly changing and boundaries with other online platforms and traditional metrics are fuzzy. Many online platforms cannot be easily classified and more traditional metrics, such as downloads and mentions in policy documents, have been referred to as altmetrics due to data provider policies. This review focuses on platforms, along with derived metrics, which have a clear focus on social functions; that is, those platforms which allow users to connect and interact with each other; create and reuse content; and comment on, like, and share user-provided content. Based on the reviewed literature regarding the use of social media in academia and derived metrics, we group platforms according to their major functionalities into the following categories: social networking; social bookmarking and reference management; social data sharing; video, blogging; microblogging; wikis; and social recommending, rating, and reviewing services.

The review consists of two main parts: the first examines the use of social media in academia, discusses the various functions these platforms have in the scholarly communication process, and reviews the factors that affect this use. This section is further divided into social media use by researchers, which covers the majority of studies, and by institutions and organizations. The second part focuses on research indicators based on social media activity—that is, altmetrics. It discusses the various interpretations of altmetrics, presents data collection and methodological limitations, and surveys the various altmetrics (i.e., social media metrics), according to the platforms from which they are gathered. The review ends with a discussion of the contemporary nature of scholarly communication in light of changes wrought by social media, and the implications for knowledge production and research assessment.

---

[1] Social media comprises a variety of online tools and platforms that enable users to generate content and interact with each other and has been defined as "a group of Internet-based applications that build on the ideological and technological foundations of Web 2.0, and that allow the creation and exchange of User Generated Content" (Kaplan & Haenlein, 2010, p. 61).

## 2  Social media use in academia

Social media has been adopted quickly and nearly ubiquitously across many spheres of society. However, these online platforms are not without critics, particularly for use in professional capacities: in a survey of dissemination methods, health policy researchers rated social media as the poorest dissemination method, describing it as being "incompatible with research, of high risk professionally, of uncertain efficacy, and an unfamiliar technology that they did not know how to use" (Grande et al., 2014, p. 1278). ResearchGate—a popular academic social networking site—has been called a "source of stress" (Van Noorden, 2014, p. 128), Facebook was referred to as having "zero credibility" (Van Noorden, 2014, p. 129), and blogs have been described as waste of time due to the lack of peer review (Harley, Acord, Earl-Novell, Lawrence, & King, 2010). Despite these concerns, there is considerable evidence of the increased use of social media for scholarly communication purposes (Moran, Seaman, & Tinti-Kane, 2011; e.g., Ponte & Simon, 2011; Rousidis, Garoufallou, & Balatsoukas, 2013; Tenopir et al., 2013) and acceptance of these new forms of diffusion (Piwowar, 2013b; Viney, 2013; Wilsdon et al., 2015).

Numerous surveys, interviews, and focus groups have been conducted to capture the scope of use and perceptions of social media use by scholars. Such studies mostly rely on self-reported personal use (and, to a lesser extent, institutional use) for professional purposes, as well as their attitudes towards its usefulness. However, these studies frequently utilize non-probabilistic sampling methods, and therefore the results should be generalized with caution. There are also a number of non-obtrusive studies—e.g., based on the collection of tweets—that have examined the extent and type of use by scholars of various demographic characteristics; these also have certain sampling biases that must be taken into account in interpretation. The following sections reviews this literature, examining how scholars and organizations are using social media platforms, and the factors that affect this use.

### 2.1  Scholarly use by researchers

Use of social media platforms for by researchers is high—ranging from 75 to 80% in large-scale surveys (Rowlands et al., 2011; Tenopir et al., 2013; Van Eperen & Marincola, 2011). However, the definition of social media used in these surveys is quite broad—e.g., including video conferencing (e.g., Skype) and collaborative writing tools (e.g., Google Docs). Given that use of a single platform varies considerably—e.g., less than 10% of scholars reported using Twitter (Rowlands et al., 2011), while 46% used ResearchGate (Van Noorden, 2014), and more than 55% used YouTube (Tenopir et al., 2013)—it is necessary to discuss the use of various types of social media separately. Furthermore, there is a distinction among types of use, with studies showing higher uses of social media for dissemination, consumption, communication, and promotion (e.g., Arcila-Calderón, Piñuel-Raigada, & Calderín-Cruz, 2013; Van Noorden, 2014), and fewer instances of use for creation (i.e., using social media to construct scholarship) (British Library et al., 2012; Carpenter, Wetheridge, Tanner, & Smith, 2012; Procter et al., 2010b; Tenopir et al., 2013).

**Social networking**

Social networking sites are "web-based services that allow individuals to (1) construct a public or semi-public profile within a bounded system, (2) articulate a list of other users with whom they share a connection, and (3) view and traverse their list of connections and those made by others within the system" (boyd & Ellison, 2007, p. 211). These sites emerged in the late 1990s and several have been founded (and have folded) since that time (boyd & Ellison, 2007). A variety of social networking sites are used for scholarly communication purposes: from those aimed at the general public (e.g., Facebook, LinkedIn, Google+) to sites targeted at scholars (e.g., ResearchGate, Academia.edu, VIVO).

Among these sites, Facebook has repeatedly been show as the most frequently used platform (Bowman, 2015; Capano, Deris, & Desjardins, 2009; Haustein, Peters, Bar-Ilan, et al., 2014; Madhusudhan, 2012; Moran et al., 2011)—perhaps unsurprisingly given that Facebook boasts 1.5 billion users, representing half of the world's online population (Hope, 2015). LinkedIn is also a highly used social networking site for academics (Bowman, 2015; Haustein, Peters, Bar-Ilan, et al., 2014; Loeb et al., 2014; Wilson & Starkweather, 2014). However, the percentage of scholars who use social networking for professional purposes is much lower than those who use it for personal reasons (Loeb et al., 2014; Nentwich & König, 2014, p. 114). When investigating only professional use, approximately a quarter of respondents reported using Facebook (Procter et al., 2010), LinkedIn (Mas-Bleda, Thelwall, Kousha, & Aguillo, 2014), and ResearchGate (Bowman, 2015), with lower rates reported for Academia.edu (Bowman, 2015; Mas-Bleda et al., 2014). However, rates of use reported vary by platform (Van Noorden, 2014) and by geographic region (Ortega, 2015).

Scholars report that one of the barriers for professional use is the trade-off between the amount of time it takes to learn a new tool and the expected advantages (Dantonio, Makri, & Blandford, 2012; Davis, Coppock, & Vowell, 2012). Connecting with other researchers (for community building or collaboration), disseminating research, and following the research output of others are primary motivations for scholarly use of social networking sites (Arcila-Calderón et al., 2013; Nández & Borrego, 2013; Veletsianos, 2012), though motivations differ by user group (e.g., students, post-docs, and lecturers) (British Library et al., 2012; Nández & Borrego, 2013) and by domain area (Chakraborty, 2012; Elsayed, 2015). Academic social networking sites are also used for professional branding—in a study of Academia.edu, the majority of respondents reported that the profile functioned like an online business card (Jordan, 2014; Van Noorden, 2014), although many profiles contain little information (Thelwall & Kousha, 2014). Impression management is a growing concern on social media platforms, as scholars gravitate to these sites to construct and display their scholarly identity (Bowman, 2015; Veletsianos, 2012, 2013) and accrue academic capital (Thelwall & Kousha, 2014, 2015). The tensions that this creates in terms of a professional and personal identity has been discussed in a number of studies (e.g., Bowman, 2015; Veletsianos & Kimmons, 2013), particularly in light of the widely-publicized cases of censuring of academics for lack of civility on social media (e.g., Bennett-Smith, 2013; Berrett, 2010; Herman, 2014; McMurtrie, 2015; Rothschild & Unglesbee, 2013).

**Social bookmarking and reference management**

Bookmarking and reference managers allow users to favorite or save publications, organize bibliographic materials, and share research with others. All genre types can be bookmarked and managed, though journal articles are the most widely used type (Borrego & Fry, 2012). Frequently mentioned social platforms in scholarly communication research include research-specific tools such as Mendeley, Zotero, CiteULike, BibSonomy, and Connotea (now defunct) as well as general tools such as Delicious and Digg (Hammond, Hannay, Lund, & Scott, 2005; Hull, Pettifer, & Kell, 2008; Priem & Hemminger, 2010; Reher & Haustein, 2010). Users can leave comments, rate papers, create their own tags, and (with some platforms) cite entries in their own documents. Reference managers often have built-in social networking components as well, where users can join groups, share documents, and follow other users. These integrated platforms have been termed *academic social networking services* (Jeng, He, & Jiang, 2015), blending socially-oriented reference management systems with scientifically-oriented social networking sites.

Mendeley has been identified as a particularly promising source of data for altmetrics (e.g., Costas, Zahedi, & Wouters, 2015). However, its adoption has been relatively low: surveys of various subpopulations have found use rates at less than 10% (Bowman, 2015; Mas-Bleda et al., 2014; Ortega, 2015; Van Noorden, 2014), though more than one-quarter of bibliometricians report Mendeley use (Haustein, Peters, Bar-Ilan, et al., 2014). Similarly low rates have been found for other social reference managers (Haustein, Peters, Bar-Ilan, et al., 2014; Pscheida, Albrecht, Herbst, Minet, & Köhler, 2013).

Doctoral students and junior researchers are the largest reader group in Mendeley (Haustein & Larivière, 2014; Jeng et al., 2015; Zahedi, Costas, & Wouters, 2014a). This is an important distinction as significant differences in Mendeley use have been found by status: faculty members are more likely to use it to publicize their publications, while students are more likely to search for publications (Mohammadi, 2014; Mohammadi, Thelwall, & Kousha, 2015). Groups are also used in particularistic ways. Initial research suggests that Mendeley groups could provide opportunities for interdisciplinary research (Oh & Jeng, 2011), though discipline appears to be an important variable in use— Mohammadi and Thelwall (2014) found that the discipline of papers in the social sciences and humanities aligned mostly with the disciplines of the readers. Gunn (2013) found stronger correlations between readers and citations within the same discipline than from across multiple disciplines. Behavioral factors also heavily influence the success of various groups (Jeng, He, Jiang, & Zhang, 2012). However, the social networking mechanisms of bookmarking and reference management platforms are used far less frequently than the research-based features (Jeng et al., 2015; Jordan, 2014; Mohammadi, 2014; Mohammadi, Thelwall, & Kousha, 2015).

**Social data sharing**

Data sharing has become a requirement of several funders and journals (Piwowar & Chapman, 2010; Tenopir et al., 2011) on the basis of enhanced verifiability and replicability in science. In this context, a number of data sharing platforms have been established, many of which focus on specific fields or individual communities (Costas, Meijer, Zahedi, & Wouters, 2013). Social data sharing platforms provide an infrastructure to share various types of scholarly objects—including datasets, software code, figures, presentation slides and videos—and for users to interact with these objects (e.g., comment on, favorite, like, and reuse). Platforms such as Figshare and SlideShare disseminate scholars' various types of research outputs such as datasets, figures, infographics, documents, videos, posters, or presentation slides (Enis, 2013) and displays views, likes, and shares by other users (Mas-Bleda et al., 2014). GitHub provides for uploading and storing of software code, which allows users to modify and expand existing code (Dabbish, Stuart, Tsay, & Herbsleb, 2012), which has been shown to lead to enhanced collaboration among developers (Thung, Bissyande, Lo, & Jiang, 2013). As with other social data sharing platforms, usage statistics on the number of view and contributions to a project are provided (Kubilius, 2014). The registry of research data repositories, re3data.org, has indexed more than 1,200 as of May 2015[2]. However, only a few of these repositories (i.e., Figshare, SlideShare and Github) include social functionalities and have reached a certain level of participation from scholars (e.g., Begel, Bosch, & Storey, 2013; Kubilius, 2014).

Data sharing and reuse on social media are not yet common, and surveys on the adoption and use of such platforms are rare. One study (Van Noorden, 2014) found that less than 10% of surveyed researchers were aware of Figshare and as few as 0.5% visited the site regularly. A study of SlideShare found that less than 5% of highly-cited researchers had uploads (Mas-Bleda et al., 2014), and that these uploads had relatively few views, likes, or shares. Furthermore, other studies (e.g., Peters, Kraker, Lex, Gumpenberger, & Gorraiz, 2015; Torres-Salinas, Jiménez-Contreras, & Robinson-García, 2014) suggest that data are rarely cited.

---

[2] http://www.re3data.org/2015/05/datacite-to-manage-and-develop-re3data-org/

**Video**

Video provides yet another genre for social interaction and scholarly communication (Kousha, Thelwall, & Abdoli, 2012; Sugimoto & Thelwall, 2013). Of the various video sharing platforms, YouTube, launched in 2005, is by far the most popular (Thelwall, Kousha, Weller, & Puschmann, 2012). Although the platform provides a broad spectrum of content, videos in the Science & Technology category are prominent among highly discussed videos (Thelwall, Sud, & Vis, 2012) and scholars increasingly cite YouTube videos in published research, although citation rates remain very low (Kousha et al., 2012). TED Talks—videos from TED conferences found online—are more focused on science and technology and have been shown to be as one of the most successful contemporary scholarly communication initiatives (Sugimoto et al., 2013; Sugimoto & Thelwall, 2013), though academics are a minority among TED presenters (Sugimoto et al., 2013). Commenting functionality is available, both through the TED website and on the targeted YouTube channel, allowing for an analysis of audience reception to the videos (Tsou, Thelwall, Mongeon, & Sugimoto, 2014). A study of UK scholars reports that the majority of respondents engaged with video for scholarly communication purposes (Tenopir et al., 2013), yet only 20% have ever created in that genre. Among British PhD students, 17% had used videos and podcasts passively for research, while 8% had actively contributed (British Library et al., 2012). This highlights the passive role of scholars in video sharing, common across social media platforms.

**Blogging**

Blogs began in the mid-1990s and were considered ubiquitous by the mid-2000s (Gillmor, 2006; Hank, 2011; Lenhart & Fox, 2006; Rainie, 2005). Scholarly blogs emerged during this time with their own neologisms (e.g., blogademia, blawgosphere, bloggership) and body of research (Hank, 2011) and were considered to change the exclusive structure of scholarly communication (Mortensen & Walker, 2002). While most scholarly bloggers utilized a standard blog service provider (e.g., Live Journal, WordPress), aggregation of scholarly blogs onto platforms or directories was less systematic. Furthermore, scholarly blogs vary in format and content, making it difficult to constitute pure sampling frames or analyze behaviors on scholarly blogs in homogeneous ways (Davies & Merchant, 2007; Kouper, 2010; Walker, 2006). Several sources—for example, the Academic Blog Portal, ScienceBlogs, Law Professor Blog network—attempted to collect scholarly blogs, but they were rarely comprehensive and quickly dated (Hank, 2011). The dynamic nature of this landscape caused concerns for those attempting to describe the frequency or nature of use of blogs for scholarly communication (Hank, 2011). For example, Technorati, considered to be one of the largest index of blogs, deleted their entire blog directory in 2014[3]. Individual blogs are also subject to abrupt cancellations and deletions, making questionable the degree to which blogging meets the *permanence* criteria of scholarly communication (Hank, 2011).

ResearchBlogging.org (RB)— "an aggregator of blog posts referencing peer-reviewed research in a structured manner" (Shema, Bar-Ilan, & Thelwall, 2015, p. 3)—was launched in 2007 and has been a fairly stable structure in the scholarly blogging environment. RB both aggregates and—through the use of the RB icon—*credentials* scholarly blogs (Shema et al., 2015). This makes the platform ripe for empirical study, which have demonstrated, among other things, that scholarly blogs tend to link to more recent literature than journal articles (Groth & Gurney, 2010; Shema et al., 2015) and refer most frequently to prominent journals (i.e., *Nature*, *PLOS ONE*, *PNAS*, and *Science*) (Fausto et al., 2012; Groth & Gurney, 2010; Shema, Bar-Ilan, & Thelwall, 2012, 2014; Shema et al., 2015).

---

[3] http://www.business2community.com/social-media/technorati-worlds-largest-blog-directory-gone-0915716

Many studies reinforce the review function of blogs—that is, to disseminate, comment on, or critique published research (Bonetta, 2007; Kjellberg, 2010; Mahrt & Puschmann, 2014; Puschmann, 2014; Wilkins, 2008). Blogs are also frequent sources of "academic culture critique" (Mewburn & Thomson, 2013, p. 1110)—that is, platforms upon which scholars can critically discuss the mechanisms underlying academic systems (Shema et al., 2015). The informality of the genre (Mewburn & Thomson, 2013) and the ability to circumvent traditional publishing barriers has led advocates to claim that blogging can invert traditional academic power hierarchies (Walker, 2006), allowing people to construct scholarly identities outside of formal institutionalization (Ewins, 2005; Luzón, 2011; Potter, 2012) and democratize the scientific system (Gijón, 2013).

Another positive characteristic of blogs is their "inherently social" nature (Walker, 2006, p. 132) (see also Kjellberg, 2010; Luzón, 2011). Scholars have noted the potential for "communal scholarship" (Hendrick, 2012) made by linking and commenting, calling the platform "a new 'third place' for academic discourse" (Halavais, 2006, p. 117). Commenting functionalities were seen as making possible the "shift from public understanding to public engagement with science" (Kouper, 2010, p. 1). The comments have been mined to quantify the relationship between the procedures and consumers of blogs, by comparing the linguistic properties of blog posts with reader comments (Mahrt & Puschmann, 2014). Text analysis has also been applied to posts (for a better understanding of the nature of the communication see Luzón (2012) and Puschmann and Bastos (2015)) and to links to understand the construction of communities (Luzón, 2009).

The *tension* between traditional/institutional and alternative/democratic means of research dissemination is often evoked in discussions of scholarly blogging (see, e.g., (Hendricks, 2010). Although the use of pseudonyms was popular in early blogging, it is no longer the norm. Most authors use their real name (Kovic, Lulic, & Brumini, 2008; Shema et al., 2012, 2015), which has allowed researchers to classify bloggers. Unsurprisingly, the large majority of research bloggers hold advanced degrees and are affiliated with academic institutions (Kouper, 2010; Kovic et al., 2008; Puschmann & Mahrt, 2012; Shema et al., 2012). Studies have also provided evidence of high rates of blogging among certain subpopulations: for example, approximately one-third of German university staff (Pscheida et al., 2013) and one fifth of UK doctoral students use blogs (Carpenter et al., 2012).

Academics are not only producers, but also consumers of blogs: a 2007 survey of medical bloggers found that the large majority (86%) read blogs to find medical news (Kovic et al., 2008), academic blogging communities have been found to exhibit small-world network properties (e.g., Wang, Jiang, & Ma, 2010), and a higher proportion of British doctoral students indicated that they followed blogs (23%) compared with those who contributed to them (9%) (Carpenter et al., 2012). This may work at odds with the perceived audience of blogs: in a study of SciLogs, the overwhelming majority of bloggers considered the general public to be the main audience—at 80%, compared to the 40% who reported colleagues and students as the main audience (Puschmann & Mahrt, 2012). This suggests that, at least for this platform, bloggers conceived their work as science popularization rather than scholarly communication. This was a critical characteristic explored in Mahrt and Puschmann (2014), who defined science blogging as "the use of blogs for science communication" (p. 1). It has been similarly likened to a space for public intellectualism (Kirkup, 2010; Walker, 2006) and as a form of activism to combat perceived biased or pseudoscience (Riesch & Mendel, 2014). Yet, there remains a tension between science bloggers and science journalists, with many science journals dismissing the value of science blogs (Colson, 2011).

Legitimization of the blog genre has come in the form of blogging by both respected brands (e.g., *Nature*, *Wired*, *PLOS*, *The Guardian*) and prominent individuals (e.g., Krugman, Gowers, and Tao) (Shema et al., 2015), though attitudes towards the usefulness of blogs as a research source and their credibility vary. Proponents argue for the scholarliness of blogging (Harley et al., 2010; D. G. Smith, 2006). Others argue that it supplements, rather than replaces traditional publishing and serves as "an alternative form of 'peer review' that is more competitive, open, and transparent than the traditional peer review process" (Solum, 2006, p. 1088). However, despite these claims, blogging was rated as one of the least efficacious methods of disseminating research by health policy researchers, ranking only above Facebook (Grande et al., 2014). Furthermore, while there has been anecdotal evidence of the use of blogs in promotion and tenure (e.g., (Podgor, 2006) the consensus seems to suggest that most institutions do not value blogging as highly as publishing in traditional outlets, or consider blogging as a measure of service rather than research activity (Hendricks, 2010, para. 30).

Generalizing from these studies is s relatively difficult, as many of them rely on convenience samples based on listservs (e.g., Gruzd & Goertzen, 2013; Wilson & Starkweather, 2014) rather than on random samples, or are based on small samples or single case studies (e.g., Shanahan, 2011; Wade & Sharp, 2013). These sampling approaches may bias the results in favor of people who are particularly active online or have strong opinions about these technologies.

**Microblogging**

Microblogging developed out of a particular blogging practice, wherein bloggers would post small messages or single files on a blog post. Blogs that focused on such "microposts" were then termed "tumblelogs" and were described as "a quick and dirty stream of consciousness" kind of blogging (Kottke, 2005, para. 2). Separate platforms were subsequently developed to facilitate this type of posting—among the most popular microblogs are Twitter (launched in 2006), tumblr (launched in 2007), FriendFeed (launched in 2007 and available in several languages), Plurk (launched in 2008 and popular in Taiwan), and Sina Weibo (launched in 2009 and popular in China). Contemporary microblogging platforms limit posts by character length and offer several mechanisms for social networking and sharing multimedia files. Among these, Twitter is by far the most popular (e.g., Grosseck & Holotescu, 2011; Letierce, Passant, Breslin, & Decker, 2010; Thelwall, Haustein, Larivière, & Sugimoto, 2013). As with other microblogging sites, Twitter limits posts, called "tweets", to 140 characters. It also allows users to follow other users, search tweets by keywords or hashtags, and link to other media or other tweets.

Scholars report high knowledge, but lower levels of use of microblogs compared to other forms of social media: with use rates ranging from 5% to 32% (Bowman, 2015; British Library et al., 2012; Carpenter et al., 2012; Grande et al., 2014; Gu & Widén-Wulff, 2011; Procter et al., 2010a; Pscheida et al., 2013; Rowlands et al., 2011; Tenopir et al., 2013; Van Noorden, 2014), though studies rarely distinguish between personal and professional use. Microblogs are particularly well-known, but under- (Van Noorden, 2014) or passively used (British Library et al., 2012; Carpenter et al., 2012), leading to a designation as a "hype medium" (Pscheida et al., 2013).

In a manner similar to blogging, the majority of Twitter users provide their full name and identify professionally in their Twitter account descriptions (Bowman, 2015; Chretien, Azar, & Kind, 2011; Hadgu & Jäschke, 2014). Those who do have been shown to have more followers than those who do not (Lulic & Kovic, 2013). Networks of followers can shed light on those who are influential in scholarly dissemination—for example, Holmberg, Bowman, Haustein, and Peters (2014) mapped the Twitter interactions of 32 astrophysicists with other users and found that the largest group of Twitter users mentioned by astrophysicists were science communicators. There are also networking benefits to tweeting: one study demonstrated that conference speakers and participants who tweet at conferences saw statistically significant gains in followers during the conference (Sopan, Rey, Butler, & Shneiderman, 2012). However, identifying professionally might also have negative effects, as demonstrated by high-profile cases of firing or not being hired for content of social media tweets (e.g., Chretien et al., 2011; Herman, 2014; Ingeno, 2013; Rothschild & Unglesbee, 2013).

Microblogging, like other social media platforms, blurs the boundaries between professional and personal (Bowman, 2015). Although the majority of academics using Twitter regularly seem to use it professionally to some degree—for research-related discussions and communicating with others in the field (Van Noorden, 2014)—a large share of their activity is personal (Bowman, 2015; Haustein, Bowman, Holmberg, Peters, & Larivière, 2014; Mou, 2014; Pscheida et al., 2013; Van Noorden, 2014). However, classifying tweet content is difficult due to the brevity of tweets (Bowman, 2015). In addition, the proportion of scholarly posts from an account varies dramatically by individual (Chretien et al., 2011; Loeb et al., 2014; Mou, 2014; Priem, Costello, & Dzuba, 2012) and by discipline (Holmberg & Thelwall, 2014), which makes it difficult to identity accounts as "scientific", though many attempts have been made to create lists of tweeting scientists (e.g., Science Pond) (Bonetta, 2009).

Scholarly tweets tend to contain indirect links (Holmberg & Thelwall, 2014) to recent journal articles (Eysenbach, 2011; Holmberg & Thelwall, 2014; Priem & Costello, 2010). Blogs are also common destinations for these indirect links (Letierce et al., 2010; Priem & Costello, 2010; Weller, Dröge, & Puschmann, 2011; Weller & Puschmann, 2011) and direct links are more frequently employed for open access articles (Priem & Costello, 2010). Tweets linking to scholarly articles tends to be limited to the exact title of the article, retweets from the publishing journal, or slightly modified retweets, and are fairly neutral (Friedrich, Bowman, Stock, & Haustein, 2015; Thelwall, Tsou, Weingart, Holmberg, & Haustein, 2013). Hashtags, another frequently utilized affordance in microblogging (Bowman, 2015), tend to be used to indicate the general topic referred to by the tweet, rather than what the tweeted object is (Letierce et al., 2010). The use of Twitter-specific affordances such as retweets by scholars has also been analyzed (Bowman, 2015; Haustein, Bowman, Holmberg, et al., 2014; Priem & Costello, 2010).

Conference chatter is another widely studied area in the realm of scholarly microblogging. Twitter use at conferences is generally carried out by a minority of participants (Chaudhry, Glode, Gillman, & Miller, 2012; Cochran, Kao, Gusani, Suliburk, & Nwomeh, 2014; Desai et al., 2012; McKendrick, Cumming, & Lee, 2012; Mishori, Levy, & Donvan, 2014; Mishori, Singh, Levy, & Newport, 2014; Reinhardt, Ebner, Beham, & Costa, 2009; Weller et al., 2011; Weller & Puschmann, 2011). However, several conferences have seen an increase in Twitter adoption over time (Chaudhry et al., 2012; Hawkins, Duszak, & Rawson, 2014; Mishori, Levy, et al., 2014). Although most researchers report never using Twitter for outreach (Wilkinson & Weitkamp, 2013), tweeting during conferences represents a type of outreach—disseminating the content and conversation of a physical event to a virtual audience. In fact, the vast majority of Twitter users engaging in conference tweets at some conferences are not in-person attendees (Sopan et al., 2012).

Scholarly discussions on microblogging platforms are not limited to conferences. For instance, health professionals have used Twitter to organize journal clubs (e.g., Thangasamy et al., 2014), wherein practitioners read and discuss scientific papers relevant to their work. Twitter has also been used to critique and correction the scientific record. For example, data from a genomics paper was re-analyzed and the result was posted on Twitter (e.g., Woolston, 2015), in a manner arguably faster and to a wider audience than traditional publishing channels.

**Wikis**

Wikis are collaborative content management platforms enabled by web browsers and embedded markup languages. Wikipedia, launched in 2001, is a highly popular and well-established online encyclopedia that provides opportunities for crowdsourcing content (Okoli, Mehdi, Mesgari, Nielsen, & Lanamäki, 2014) and is increasingly used professionally by researchers (Moeller, 2009). Despite "vociferous critics" (Okoli et al., 2014, p. 2385), scholars tend to view of Wikipedia as a credible source (Chesney, 2006; Dooley, 2010), particularly for educational use (Aibar, Lerga, Lladós, Meseguer, & Minguillon, 2013; Taraborelli,

Mietchen, Alevizou, & Gill, 2011). It has been suggested as a good starting point for research (Hodis et al., 2008; Williams et al., 2009) and studies have found fairly high rates (i.e., 30-45%) of use for searching information (Archambault et al., 2013; Carpenter et al., 2012; Gu & Widén-Wulff, 2011), but much lower rates of contribution by scholars (Bender et al., 2011; Carpenter et al., 2012; Giles, 2005; Weller, Dornstädter, Freimanis, Klein, & Perez, 2010; Xiao & Askin, 2014).

Wikipedia has been advocated as a replacement for traditional publishing and peer review models (Xiao & Askin, 2012) and pleas have been made to encourage experts to contribute (Rush & Tracy, 2010). Despite this, contribution rates remain low—likely hindered by the lack of explicit authorship in Wikipedia, a cornerstone of the traditional academic reward system (Black, 2008; Butler, 2008; Callaway, 2010; Whitworth & Friedman, 2009). Citations to scholarly documents—another critical component in the reward system—are increasingly being found in Wikipedia entries (Bould et al., 2014; Park, 2011; Rousidis et al., 2013), but are not yet seen as valid impact indicators (Haustein, Peters, Bar-Ilan, et al., 2014).

The wiki format has also been adopted for other academic purposes. For example, wikis have been used internally for writing early drafts of articles and documenting the research process—i.e., taking the lab notebook online (Collins & Jubb, 2012; University of Edinburgh Digital Curation Centre, 2010). Scholars have also highlighted the enhanced functionality of using web-based platforms for disseminating research, such as the ability to render three-dimensional images (Hodis et al., 2008). However, existing studies on the contribution of wikis to scholarly communication tend to be rather anecdotal (for a review, see Okoli et al. (2014)), and large-scale, cross-disciplinary studies of scholarly use of this platform are lacking.

**Social recommending, rating, and reviewing**

According to the altmetrics manifesto (Priem et al., 2010, para. 1), altmetrics can serve as filters, which "reflect the broad, rapid impact of scholarship in this burgeoning ecosystem". Although the commenting and discussion function of social media platforms can indeed serve as filters, other systems have also been developed specifically for filtering the most relevant scientific content through recommendations and ratings. Among those, F1000Prime (formerly F1000) is the most popular, focusing on biological and medical publications. In this system, selected experts—currently 5,000 so-called "Faculty Members"—recommend, rate, and review the most important articles of their subfields (Li & Thelwall, 2012; Waltman & Costas, 2014). Pubpeer—an online journal club—extends this model, allowing any user to anonymously comment on scientific documents with a DOI or arXiv id (Townsend, 2013).

On these platforms, the boundary between recommending and reviewing becomes blurred. Peer review is a central part of the scholarly communication system, as it functions as a quality control and gatekeeping mechanism. While such gatekeeping has traditionally been closed, it has, over the recent years, become more transparent and visible online: open reviews can be signed, disclosed, editor-mediated, transparent or crowdsourced and happen prior to, after, or synchronous with publication (Ford, 2013; Lee, Sugimoto, Zhang, & Cronin, 2013; Tattersall, 2015). Many of the early post-publication approaches, including online commenting on journal platforms, were relatively unsuccessful. For instance, the *British Medical Journal* (BMJ) started an early trial in 1999 (R. Smith, 1999), and found no significant differences in review quality, decision, or time to completion. However, referees were more likely to decline to review (van Rooyen, Godlee, Evans, Black, & Smith, 1999), which suggests that complete transparency might not be acceptable to all. *Nature* held a four-month trial phase of open peer review wherein authors were asked if they would allow their papers to be open to technical commenting online, in addition to the regular peer review process (Anonymous, 2006). Only 5% of authors agreed, and only about half of these received any comments. The option to open papers up to comment was subsequently abandoned. Similarly low rates of post-publication commenting have been found on *PLOS ONE*, with the majority of the comments written by authors and editors (Adie, 2009). This could be anticipated given the results of perceptions of open peer review, which has been consistently rated poorly by scholars (see Lee, Sugimoto, Zhang, and Cronin (2013) for a review). Despite these early failed attempts, there seems to be increasing interest in open peer review from new journals such as F1000Research (Swoger, 2013) and PeerJ (Binfield, 2013). Platforms have also appeared to incentivize reviewing activity across journals, such as Publons, which specializes in crediting scholars for their gatekeeping activity (Gasparyan, Gerasimov, Voronov, & Kitas, 2015).

There are also a host of platforms which are being used informally to discuss and rate scholarly material. Reddit, for example, is a general topic platform where users can submit, discuss and rate online content. Historically, mentions of scientific journals on Reddit have been rare (Thelwall, Haustein, et al., 2013). However, several new subreddits—e.g., science subreddit[4], Ask Me Anything sessions[5]--have recently been launched, focusing on the discussion of scientific information. Sites like Amazon (Kousha & Thelwall, 2015) and Goodreads (Zuccala, Verleysen, Cornacchia, & Engels, 2015), which allow users to comment on and rate books, has also been mined as potential source for the compilation of impact indicators.

## 2.2 Scholarly use by institutions and organizations

Scholarly institutions—such as universities, libraries, professional societies, and publishers—are increasingly using social media platforms for diffusing and promoting research. While such tools are also used by universities to attract new students (Greenwood, 2012; Hayes, Ruschman, & Walker, 2009; Nyangau & Bado, 2012)—as well as by academic libraries to promote their services (Boateng & Quan Liu, 2014; Hussain, 2015; Zohoorian-Fooladi & Abrizah, 2013)—such marketing use will not be covered in this section. Instead, it will focus on how higher education institutions, journals, publishers, and even pharmaceutical corporations (Feeny, Magee, & Wilson, 2014), are using social media tools in a scholarly communication context.

**Higher education**

Higher education institutions have been shown to use social media for disseminating their research to a local and wider community—with variations according to the size and structure of the institutions (Forkosh-Baruch & Hershkovitz, 2012; Prabhu & Rosenkrantz, 2015). The pervasive use of social networking sites by faculty members and students has also led to numerous studies on their integration in the pedagogical mission of higher education institutions (e.g., Arnold & Paulus, 2010; Brown & Green, 2010; Kabilan, Ahmad, & Abidin, 2010; Kalashyan et al., 2013; Klein, Niebuhr, & D'Alessandro, 2013), as well as on the tensions that arise when faculty and students interact on these platforms (Hank, Tsou, Sugimoto, & Pomerantz, 2014; Sugimoto, Hank, Bowman, & Pomerantz, 2015). These tensions are exacerbated by the lack of social media policies of higher education institutions (Pomerantz, Hank, & Sugimoto, 2015). The policies in existence focus on regulations for those formally employed in communication and marketing efforts (Pomerantz et al., 2015), rather than prescriptions for use by members of the community, and guidelines vary by geographic region (Pasquini & Evangelopoulos, 2015).

**Libraries**

---

[4] https://www.reddit.com/r/science/

[5] http://blogs.plos.org/plos/2015/06/update-on-plos-science-wednesday-redditscience-ama-series-upcoming-featured-plos-authors/

In addition to creating institutional accounts on platforms such as Twitter and Facebook, libraries provide services to support researchers' use of social media tools and metrics (Lapinski, Piwowar, & Priem, 2013; Rodgers & Barbrow, 2013; Roemer & Borchardt, 2013) and are making use of social media products tailored specifically to libraries (Rodgers & Barbrow, 2013). One example is Mendeley Institutional Edition, which mines Mendeley documents, annotations, and behavior and provides these data to libraries (Galligan & Dyas-Correia, 2013). Libraries can use them for collection management, in a manner similar to other usage data, such as COUNTER statistics (Galligan & Dyas-Correia, 2013). Konkiel and Scherer (2013) suggest complementing traditional usage statistics with social media visibility for repository content. Although there is wide acceptance and use of social media tools, concerns are beginning to be raised regarding patron privacy (Lamdan, 2015).

**Journals**

Journals have also increasingly adopted social media tools, though levels of adoption vary considerably: Kortelainen and Katvala (2012) found that 9% of journals had a Facebook account, while Kamel Boulos and Anderson (2012) found a percentage more than 8 times higher (80%). Reports of presence on Twitter range from 15& (Amir et al., 2014; Kortelainen & Katvala, 2012), to 25% (Nason et al., 2015), and 44% (Kamel Boulos & Anderson, 2012). Such variability suggests that this landscape is still highly dynamic and varies considerably across domains and sampling frame. Beside maintaining accounts, which are often used to share news and articles (Zedda & Barbaro, 2015), some journals have started to ask authors to provide so-called tweetable abstracts that journals can use to promote papers (Darling, Shiffman, Côté, & Drew, 2013). In addition to the promotion of published articles, a popular use of social media by journals, particularly in the medical sciences, is the creation of virtual journal clubs (Leung, Siassakos, & Khan, 2015; Mehta & Flickinger, 2014; Rezaie, Swaminathan, Chan, Shaikh, & Lin, 2015; Thangasamy et al., 2014; Thoma, Rolston, & Lin, 2014; Topf & Hiremath, 2015; Whitburn, Walshe, & Sleeman, 2015), combining the promotion of articles with post-publication peer review and community discussions. Some journals have also encouraged their authors to contribute actively to Wikipedia (Butler, 2008; Maskalyk, 2014) and engage in other outreach activities through social media (Micieli & Micieli, 2012).

**Publishers and professional associations**

Publishers and professional associations have sought to use social media both to disseminate information and to connect with a potentially wider audience (Zedda & Barbaro, 2015). For a subset of biomedical publishers, Zedda and Barbaro (2015) found that the large majority (nearly 90%) utilized Twitter, 80% had a Facebook account and YouTube, blogs, and podcasts were maintained by 42%, 41% and 30% of the publishers, respectively. Nature Publishing Group (NPG), which was identified as the most active publisher on social media (Zedda & Barbaro, 2015), syndicated a list of approved blogs, along with some written by its own staff, while the Public Library of Science (PLOS) has used staff and science bloggers to write about PLOS articles (Stewart, Procter, Williams, & Poschen, 2013).

The different publishing/business models of publishers affect their adoption of Web 2.0 tools: tensions arose between the business interests and technology-focused interests at NPG, while PLOS had more limited resources with which to experiment (Stewart et al., 2013). A few blogs from professional associations have had more success: *The Scholarly Kitchen*—an arm of the Society of Scholarly Publishing—is a notable example. Some publishers also encourage authors to promote their own work via social media platforms (Kelly & Delasalle, 2012), thus generating greater publicity for the journals they publish.

Conferences organizers have also taken advantage of Twitter by creating conferences-specific hashtags to structure online conversation, which allows for the creation of communities of interest and real-time conversations during events (Ferguson et al., 2014; Jalali & Wood, 2013; Weller et al., 2011). Use of Twitter for this function has led to the creation of a visualization dashboard created expressly for monitoring academic conference Twitter activity (Sopan et al., 2012). Conference tweets tend to be directly related to sessions, with a small minority focused on socializing (Chaudhry et al., 2012; Mishori, Levy, et al., 2014). Microblogging is also used as a form of note taking, where take-away points from sessions are generated (McKendrick et al., 2012).

## 2.3 Factors affecting social media use

There is wide variability in the use and acceptance of social media tools, and much of the published research has sought to identify factors of differentiation. Previously demonstrated differences in scholarly communication hold true in this new environment: specifically, differences are seen in age, academic rank, gender, discipline, country, and language in the degree to which scholars adopt and use such technology. However, given the various populations on which they are based as well as the points in time at which data was gathered, results are often contradictory.

**Age**

Perceptions of credibility of social media vary by age, with younger scholars having a more positive perception than those held by older scholars (Nicholas et al., 2014). Mixed results have been found regarding their use for scholarly communication: studies have found no difference by age (Hadgu & Jäschke, 2014; Rowlands et al., 2011), higher rates of use for older scholars (Procter et al., 2010a), and higher rates by younger scholars (Bowman, 2015; Oladejo, Adelua, & Ige, 2013). Differences have also been noted in terms of type of social media platform—for example, younger scholars are more likely to use blogs, RSSfeeds, and Twitter than other platforms (Tenopir et al., 2013). The inconclusive and conflicting nature of these results suggest that more—particularly longitudinal—studies are needed in this area.

**Academic rank and status**

Social media use has been shown to vary by rank and platform, with many studies obtaining divergent results. Mansour (2015) showed that assistant professors from Kuwait were more likely to use social media than lecturers, while Procter et al. (2010a) showed that, for the UK, social media use was higher for doctoral students and professors than for lecturers and readers. However, a study of Australian doctoral students found reluctance to use social media for research (Dowling & Wilson, 2015). It has also been noted that doctoral students are not a monolithic group in terms of social media practices and that local and disciplinary cultures may also shape practices (Coverdale, 2011). Harley et al. (2010) suggested that early career scholars are dissuaded from using new platforms because of both implicit and explicit requirements associated with academic advancement. This may be due to the pressure they encounter to follow traditional scholarly communication models (Acord & Harley, 2012) as well as concerns from junior faculty about impression management (Grande et al., 2014).

In terms of specific platforms, doctoral students have been shown to be the major user group for social bookmarking services (Haustein & Larivière, 2014; Mohammadi, 2014; Mohammadi, Thelwall, Haustein, & Larivière, 2015; Zahedi, Costas, et al., 2014a), though this may change over time these scholars advance in rank. Students and faculty have been shown to be roughly equally represented on blogs (Shema et al., 2012). A u-shaped curve has been identified in terms of Twitter use: respondents with 7 to 9 years of academic experience (roughly corresponding to time to tenure) were proportionally more likely to use Twitter than those with less than 7 or more than 10 years of academic experience (Bowman, 2015). Type of use varies as well: assistant professors were, for example, more likely than associate and full professors to use social media for creating and maintaining research connections (Gruzd, Staves, & Wilk, 2012). Junior scholars were also more likely to believe that social media are useful for disseminating research (Grande et al., 2014).

**Gender**

Gender disparities in scholarly communication are well-documented (e.g., Larivière, Ni, Gingras, Cronin, & Sugimoto, 2013) and appear to persist in the social media environment. Scholarly communication on social media remains mostly male-dominated, despite the generally high use of general social media tools by women (Oladejo et al., 2013; Procter et al., 2010a). This is particularly true in the case of blog authorship, which has been repeatedly shown to be male-dominated (Kovic et al., 2008; Mahrt & Puschmann, 2014; Puschmann & Mahrt, 2012; Shema et al., 2012). Studies of gender and microblogging provide mixed results, with some demonstrating higher male use (Birkholz, Seeber, & Holmberg, 2015; Tsou et al., 2015) and another study showing no difference (Bowman, 2015).

Evidence of gendered use of the platforms is scarce. Looking broadly at various types of social media, Tenopir, Volentile, and King (2013) found no relationship between gender and content creation. However, in a study of Chinese microbloggers, women were more likely to talk about personal matters than men (Mou, 2014). Gender is also a factor in how scholars are perceived by the broader audience, particularly when dealing with audiovisual material. For example, TED talk comments contained more sentiment when the presenter was female (Tsou et al., 2014).

**Discipline**

There is a lack of consensus around disciplinary differences in scholarly social media use. Many studies have suggested field-specific hurdles to the acceptance of new forms of publishing (Acord & Harley, 2012; Cheverie, Boettcher, & Buschman, 2009) and use of social media platforms (Collins, Bulger, & Meyer, 2012; Holmberg & Thelwall, 2014). Higher use of social media platforms has been demonstrated for the biomedical sciences (Fausto et al., 2012), as well as Computer Science and Mathematics (Bowman, 2015; Kadriu, 2013; Procter et al., 2010a). However, other studies have shown research-oriented social media use—particularly for doctoral students—to be higher for those in the arts, humanities, and social sciences than for students in the sciences (British Library et al., 2012; Carpenter et al., 2012). In a study of Mendeley readership, articles in the humanities had the lowest number of readers, followed by STEM fields, with the social sciences having the highest coverage in terms of readership (Mohammadi & Thelwall, 2014; Mohammadi, Thelwall, Haustein, et al., 2015). Still other studies found no disciplinary differences in the use of certain social media platforms (Priem, Costello, et al., 2012) or the creation of new content (Tenopir et al., 2013). Given the various field delineations as well as the different populations on which these analyses are based, this lack of consensus is expected.

**Country and language**

Social media tools have been heralded as great democratizers of knowledge. However, results show that it often reproduces the same visibility disparities seen in traditional publishing and metrics (e.g., Cronin & Sugimoto, 2015). For example, English is the dominant language of scholarly blogs (Fausto et al., 2012), publications from emerging countries tend to have lower visibility on social networking platforms (Alperin, 2014, 2015) and a bias towards North American journals and reviewers could be found on F1000 (Wardle, 2010). Use of the platforms also varies: use of ResearchGate was disproportionately high in Brazil and India, and lower in China, South Korea, and Russia (Thelwall & Kousha, 2015). Varying levels of credibility of social media have also been demonstrated by country (Nicholas et al., 2014). Finally, the access to tools (e.g., lack of availability of Twitter in China and Iran) must be taken into consideration in terms of the geopolitical economy of social media tools in scholarly communication.

# 3 Social media and research evaluation

Governments and funding organizations are increasingly asking scholars to demonstrate societal impact and relevance, in addition to scientific excellence (Dinsmore, Allen, & Dolby, 2014; Higher Education Funding Council for England, 2011; Piwowar, 2013b; Viney, 2013; Wilsdon et al., 2015) . Altmetrics—coined in a tweet (Priem, 2010) and promoted through a manifesto (Priem et al., 2010)—has been advocated as a potential indicator of such impact. Of course, the idea of broadening indicators of scientific productivity and impact by capturing traces —"polymorphous mentions" (Cronin, Snyder, Rosenbaum, Martinson, & Callahan, 1998)—on the web, were not new when the term altmetrics was introduced in 2010 (Priem, 2014; Priem et al., 2010). However, it gave an umbrella term for metrics that addressed the "law of requisite variety" in scholarly communication: that is, providing a "battery of metrics to capture the full range of effects attributable to a scholar's thinking and research over time" (Cronin, 2013, p. 1091).

Several criticisms have been made of the use of altmetrics for research evaluation. Some authors have focused on the lack of validation of the metrics and limitations of data collection (e.g., "Alternative metrics," 2012; Wouters & Costas, 2012), while others have argued that altmetrics are not *impact* indicators, but rather indicators of *attention* and popularity (Crotty, 2014; Gruber, 2014; Sugimoto, 2015). Such criticism is largely due to the lack of a clear conceptual or theoretical framework for altmetrics, which would provide an interpretative lens through which motivations behind social media acts could be understood. This section aims to provide an overview of the various conceptualizations of altmetrics, the limitations of data collection, as well as of the literature on the specific types of research-related metrics compiled on the different platforms.

## 3.1 Conceptualization and classification of altmetrics

Priem defined altmetrics as the "study and use of scholarly impact measures based on activity in online tools and environments" (2014, p. 266) (Priem, Groth, & Taraborelli, 2012). This broad definition incorporates the changing landscape of tools and data aggregators and locates altmetrics as a subset of webometrics (Priem, 2014). Altmetrics has also been defined as "the creation and study of new metrics based on the Social Web for analyzing and informing scholarship"—emphasizing the social aspect of altmetric data (Adie & Roe, 2013). Although the term has seen widespread adoption, scholars have not equally embraced it. For example, the "alt" in the term has come under frequent criticism (Rousseau & Ye, 2013), as most studies show that such indicators can be considered as complementary rather than *alternative* to citations. This has led some to collapse all types of metrics under the umbrella term *scholarly metrics* (Haustein, Sugimoto, & Larivière, 2015) and altmetrics under the umbrella term *social media metrics* (Haustein, Larivière, Thelwall, Amyot, & Peters, 2014), rather than defining them in opposition to each other (e.g., traditional vs. alternative).

The heterogeneity of the data collected by the various providers aggravates these issues. For instance, different aggregators collect different sources, with Altmetric.com including mentions in policy documents (Liu, 2014) and Plum Analytics library holdings (Parkhill, 2013). Furthermore, some argue that citations can be altmetrics, when they point to non-traditional research objects, such as data (Piwowar, 2013a). There is also a common conflation between altmetrics and article-level metrics (Fenner, 2013). However, while heterogeneity creates problems, advocates argue that this also generates opportunities for multi-faceted measurements of scholarly impact. For example, Adie (2014, p. 349) notes that these metrics provide "new ways of approaching, measuring and providing evidence for impact" and Crotty emphasizes that, rather than replacing or supplementing traditional indicators, altmetrics provides "different approaches to different questions" (2014, p. 145).

The multidimensional nature of altmetrics has led data aggregators and researchers to classify the various types of altmetrics. For example, PLOS groups altmetrics based on the type of action performed, such as *viewed*, *saved*, *discussed*, *cited* and *recommended* (Lin & Fenner, 2013a), while other sources have sought to disambiguate the type of audience (e.g., scholarly vs. general public) (Piwowar, 2012). Building upon these distinctions, Haustein, Bowman, and Costas (2016) introduced a theoretical framework for social media metrics that categorizes acts upon research objects (including all forms of scholarly output) by intention (i.e., *access*, *appraise,* and *apply*) and agent (e.g., scholars, funding agencies, etc.).

## 3.2 Data collection and methodological limitations

A variety of tools have been developed to compile metrics based on social media events. Altmetric.com, Plum Analytics, Impact Story (formerly Total-Impact), and PLOS Article-Level Metrics are among the most frequently mentioned data collectors and aggregators. Numerous reviews of these tools and their strengths and weaknesses have been published (Bornmann, 2014a; Brigham, 2014; Chamberlain, 2013; Das & Mishra, 2014; Galligan & Dyas-Correia, 2013; Galloway, Pease, & Rauh, 2013; Gunn, 2014; Jobmann et al., 2014; Kwok, 2012; Melero, 2015; Neylon, Willmers, & King, 2014; Neylon et al., 2014; Neylon & Wu, 2009; Priem, 2014; Priem & Hemminger, 2010; Rasmussen & Andersen, 2013; Robinson-García, Torres-Salinas, Zahedi, & Costas, 2014; Roemer & Borchardt, 2012; Sweet, 2014; Wouters & Costas, 2012; Zahedi, Fenner, & Costas, 2014). Most of these reviews highlighted issues associated with data collection and quality, as well as with the transparency of the process.

One of the critical issues is that these aggregators concentrate on documents that have a unique object identifier, which inevitably neglects certain document types (Liu & Adie, 2013; Neylon et al., 2014). For example, Altmetric.com—arguably the most prominent altmetrics aggregator—focuses its data collection on DOIs, which has led to a de facto reduction of altmetrics studies to journal articles, excluding many types of documents and journals (Haustein, Sugimoto, et al., 2015; Taylor, 2013b) as well as most second-order event, such as the discussion of an article in a blog post or newspaper articles (Taylor, 2013a). Furthermore, even though the presence of DOIs is constantly increasing, it remains low in journals in the social sciences and humanities (Haustein, Costas, & Larivière, 2015) and from developing countries (Alperin, 2013). Data collection is even more difficult for platforms lacking APIs, such as Academia.edu and ResearchGate (Wilsdon et al., 2015).

The quality of altmetric data is further undermined by the variations that occur between curated datasets and data mined directly from APIs (Taraborelli, 2008; Wouters et al., 2015). Such idiosyncrasies make comparisons across datasets difficult (Jobmann et al., 2014; Torres-Salinas, Cabezas-Clavijo, & Jiménez-Contreras, 2013; Zahedi, Fenner, et al., 2014). For example, a comparison of Mendeley and WoS metadata showed important differences, thus affecting the retrieval of Mendeley reader counts (Zahedi, Bowman, & Haustein, 2014). Along these lines, Bar-Ilan (2014) found large fluctuations between reader counts over time, questioning their reliability. Large discrepancies between social media counts were also observed comparing results obtained through different aggregators (Jobmann et al., 2014; Zahedi, Fenner, et al., 2014). Some of these differences are due to different retrieval strategies: for example, Altmetric.com tracks only public Facebook wall posts, while PLOS ALM considers also private posts, shares, and likes (Zahedi, Fenner, et al., 2014). Moreover, Altmetric.com only collects Mendeley reader counts for documents for which a signal was obtained on other altmetrics indicators, which explains its lower Mendeley coverage (Knight, 2014 ; Robinson-García et al., 2014). These concerns motivated the creation of an altmetric working group of the National Information Standards Organization (NISO), which has sought to improve data quality standards for altmetrics (Gunn, 2014).

A few studies have questioned the opacity those behind social media acts—that is, the characteristics of those who are tweeting, liking, saving, and conducting other actions that translate into indicators. One concern is the presence of automated accounts—i.e., "bots"—and the influence of these on various indicators (Haustein, Bowman, Holmberg, et al., 2016). Another concern is the lack of information on user's accounts, which makes demographic analyses (Desai, Patwardhan, & Coore, 2014; Tsou et al., 2015) and appropriate sampling (Liang & Fu, 2015) difficult. For example, tweets have been claimed to reflect interest by the general public, although most tweets to scientific papers are more likely to come from researchers (Birkholz et al., 2015; Tsou et al., 2015). Although tools such as Altmetric.com and ImpactStory categorize the accounts generating online attention according to user types, these categories have been shown to be highly problematic (Tsou et al., 2015).

Altmetrics have often been compared with well-established citation-based metrics. Such analyses attempts to find high correlation and thereby validate the novel metrics against a gold standard or to find dissimilarity in order to argue for their alternative nature (Fausto et al., 2012; Priem, 2014). In the absence of a strong correlation between the two, social media metrics have been touted as complements to traditional metrics, rather than alternatives (Cress, 2014; Haustein, Costas, & Lariviere, 2015). However, there are many characteristics that altmetrics share with citations—such as skewness—such that the altmetrics can import many methodological solutions. For example, studies have emphasized the importance of normalization, given wide differences in social media metrics by discipline and topic (Costas, Zahedi, & Wouters, 2014; Costas et al., 2015; Haustein, Costas, et al., 2015; Haustein, Peters, Sugimoto, Thelwall, & Larivière, 2014).

## 3.3 Social media metrics

As noted, a large emphasis has been made on assessing the adequacy of social media metrics for measuring research impact, with many studies reporting the relationship between these new indicators and traditional bibliometric indicators. These studies examine the extent to which articles published in journals that used DOIs are represented on various platforms, the average attention they receive, and their correlations with citations. Studies report basic measures such as *coverage*, i.e., the percentage of documents with at least one mention on a particular platform; *density*, that is the mean number of events per document including; and *intensity*, the mean number of events excluding those without mentions (Haustein, Costas, et al., 2015). Composite indicators have also been proposed, such as the Social Media index (Thoma et al., 2015), the Altmetric score (Adie & Roe, 2013), and the Journal Social Impact score (Alhoori & Furuta, 2014). However, it has also been argued that social media metrics are too heterogeneous to be collapsed into a single metric (Lin & Fenner, 2013b).Therefore, we will examine the use of social media metrics in research evaluation by platform type.

**Social networking**

Studies have found relatively low coverage of papers on social networking sites, particularly on Facebook (with rates ranging from less than 1% to around 10%, if one takes shares, comments, and likes into account) (Alperin, 2015; Costas et al., 2014, 2015; Haustein, Costas, et al., 2015; Priem, Piwowar, & Hemminger, 2012). These proportions, however, vary by domain and genre of research product with articles from biomedical and health sciences and reviews, editorials, and news articles being more popular on Facebook (Haustein, Costas, et al., 2015). Along these lines, the correlation between Facebook counts and citations has been shown to be low (around .100), with Facebook correlating more highly with other altmetrics than with bibliometric indicators, particularly with Twitter (Barthel, Tönnies, Köhncke, Siehndel, & Balke, 2015; Costas et al., 2014; Haustein, Costas, et al., 2015). ResearchGate scores have been shown to have higher correlations with citation-based ranking systems (between .200 and .483) (Thelwall & Kousha, 2015) and with other altmetrics (Ortega, 2015); this higher correlation might be due to the level of aggregation at the researcher or institutional level rather than at the paper level.

While the use of social media platform varies according to academic rank, little is known on how these ranks affect altmetric indicators. In a study of Swiss management scholars active on ResearchGate, assistant professors were found to be more central than senior faculty (Hoffmann, Lutz, & Meckel, 2015), suggesting that these platforms may destabilize traditional hierarchies.

**Social bookmarking and reference management**

Mendeley has been shown to have the highest rates of document coverage among social bookmarking services (Li, Thelwall, & Giustini, 2012; Weller & Peters, 2012) as well as among various social media (Costas et al., 2014; Haustein, Larivière, Thelwall, et al., 2014; Priem, Piwowar, et al., 2012), with 60-80% of articles having at least one reader (Alhoori & Furuta, 2014; Bar-Ilan et al., 2012; Costas et al., 2014; Hammarfelt, 2014; Haustein, Peters, Bar-Ilan, et al., 2014; Haustein & Larivière, 2014; Haustein, Larivière, Thelwall, et al., 2014; Htoo & Na, 2015; Mohammadi, Thelwall, Haustein, et al., 2015; Priem, Piwowar, et al., 2012; Thelwall & Wilson, 2015; Torres-Salinas et al., 2013; Zahedi, Costas, et al., 2014a; Zahedi & Van Eck, 2014). However, coverage has been shown to vary according to the journal (Bar-Ilan, 2012, 2014; Li et al., 2012), field (Htoo & Na, 2015; (Mohammadi & Thelwall, 2014; Mohammadi, Thelwall, Haustein, et al., 2015) and data aggregator (Knight, 2014; Robinson-García et al., 2014) studied. Rates of coverage in CiteULike (i.e., around 30%) are usually lower than those of Mendeley (Bar-Ilan et al., 2012; Priem, Piwowar, et al., 2012; Torres-Salinas et al., 2013), though CiteULike has been shown to have higher coverage than BibSonomy and Connotea (Haustein & Siebenlist, 2011). Coverage also tends to be higher for more recent publications (Bar-Ilan et al., 2012) and for high impact journals—e.g., *Nature* and *Science* (Li et al., 2012).

Correlations between citations and Mendeley reader counts range from .2-.7, though the majority of analyses tend to report around .5 or .6 (Bar-Ilan, 2013; Bar-Ilan et al., 2012; Haustein, Larivière, Thelwall, et al., 2014; Htoo & Na, 2015; Li & Thelwall, 2012; Li et al., 2012; Maflahi & Thelwall, 2015; Mohammadi, Thelwall, Haustein, et al., 2015; Sud & Thelwall, 2013; Thelwall & Sud, 2015; Thelwall & Wilson, 2015; Weller & Peters, 2012). Variations can be largely accounted for by disciplinary differences (Mohammadi, Thelwall, Haustein, et al., 2015), with particularly low correlations in the arts and humanities (Haustein, Larivière, Thelwall, et al., 2014; Mohammadi & Thelwall, 2014). Lower correlations can also be seen for samples with earlier publication dates (Schlögl, Gorraiz, Gumpenberger, Jack, & Kraker, 2013, 2014) as well as for very recent publications, due to the time needed for citations to accumulate (Thelwall & Sud, 2015).

Stronger correlations have been found between Mendeley readership and PLOS download data (Gunn, 2013) and doctoral student readership and citation counts (Haustein & Larivière, 2014; Zahedi, Costas, et al., 2014a). F1000-reviewed papers also have higher readership than non-reviewed papers (Gunn, 2013). Correlations vary by platform: correlations between CiteULike metrics and citations, for example, are lower than those obtained for Mendeley, at about .2 to .4 (Bar-Ilan et al., 2012; Htoo & Na, 2015; Li et al., 2012). Correlational analyses have also been done to examine the relationship between number of tags and citedness, but no significant relationship was found (Haustein & Peters, 2012). This may be due in part to the low level of tagging found across all social bookmarking platforms (Good, Tennis, & Wilkinson, 2009).

**Social data sharing**

Although heavily advocated, data sharing and citing are still in their nascent stages (Peters et al., 2015; 2014) and, by extension, so too are metrics based on them (Konkiel, 2013; Peters et al., 2015). It has been argued that the widespread of such indicators might motivate researchers to share and cite data (Konkiel, 2013), and scholars have introduced a variety of indicators to measure such activities (Costas et al., 2013; Ingwersen & Chavan, 2011). However, very few studies have been done on the characteristics of such indicators.

**Videos**

There have been a handful of studies focused exclusively on video sharing. There are a number of difficulties in translating video in general, and YouTube, in particular, into appropriate research evaluation metrics (Thelwall, Kousha, et al., 2012) and it has been argued that it might be more important to evaluate creation (as a productivity measure) and views (as an impact measure) rather than citations in traditional scholarly work (Thelwall, Kousha, et al., 2012). This notion was reinforced by a comprehensive study of TEDTalks, which examined both citations to these videos as well as views and presence on other social media platforms (Sugimoto & Thelwall, 2013). The study suggested that TEDTalks are poorly represented among references, but are highly viewed and used extensively for pedagogical purposes (Sugimoto & Thelwall, 2013). Furthermore, there is no citation advantage for scholars who produce TEDTalks (Sugimoto et al., 2013). No difference was found in views between academic and non-academic presenters; however, there was a significant difference in number of comments with academic presenters receiving more comments (Sugimoto et al., 2013). This suggests that commenting behavior might be a potential indicator of engagement and impact, beyond views (Sugimoto et al., 2013).

**Blogging**

In terms of research evaluation, blogging can be used to provide both a measure of research output (blog authoring) and of impact (blog citation), with the latter having been the most analyzed. As with other metrics, both coverage and correlation have been considered when comparing blog citations with citations received in papers. Coverage has been shown to be fairly low: the highest coverage rate was for PLOS papers at 7.5% (Priem, Piwowar, et al., 2012), compared to less than 2% of all Web of Science papers (Costas et al., 2014; Haustein, Costas, et al., 2015). Studies that focused on specific domains have found lower rates of coverage (Hammarfelt, 2014; Knight, 2014). There was, however, an overemphasis on retracted papers and retracted notices in blogs (Haustein, Costas, et al., 2015). As one might expect from such low coverage, correlations between blog citations and traditional citations are weak: around 0.1 and 0.2 (Costas et al., 2014; Haustein, Costas, et al., 2015; Htoo & Na, 2015; Priem, Piwowar, et al., 2012). Higher correlations (0.3) were found with mentions in online news media (Haustein, Costas, et al., 2015). However, mentions in blogs also showed higher precision than Twitter in identifying highly-cited publications (Costas et al., 2014, 2015). Despite low correlations, studies have suggested that articles which are blogged also tend to be those with above median citation counts (e.g., Shema & Bar-Ilan, 2014), though some have cautioned that the "imperfect association between classical metrics and blog citations" (Fausto et al., 2012, p. 7) is an indication of the novelty of altmetrics.

Blog citations have been associated with appearance in the popular press (Shema & Bar-Ilan, 2014), open access publishing (Fausto et al., 2012), views and downloads of the article (Allen, Stanton, Di Pietro, & Moseley, 2013; Caron, 2006), and indexing on Mendeley (Shema et al., 2012). These results of these studies are, however, mostly determined by the overrepresentation of high-impact journals on blogs (Costas et al., 2014, 2015; Fausto et al., 2012; Groth & Gurney, 2010; Shema & Bar-Ilan, 2014; Shema et al., 2012, 2015).

**Microblogging**

Despite generating more than 500 million tweets per day, Twitter remains second only to Mendeley in terms of social media activity associated with scientific papers (Costas et al., 2015; Haustein, Costas, et al., 2015). As for other altmetrics indicators, Twitter coverage varies by discipline (Costas et al., 2014; Haustein, Bowman, Macaluso, Sugimoto, & Larivière, 2014; Haustein, Peters, Sugimoto, et al., 2014) and date of publication (Haustein, Peters, Sugimoto, et al., 2014), but has been shown to be around 10-20% (Alhoori & Furuta, 2014; Andersen & Haustein, 2015; Costas et al., 2014, 2015; Hammarfelt, 2014; Haustein, Costas, et al., 2015; Haustein, Larivière, Thelwall, et al., 2014; Haustein, Peters, Sugimoto, et al., 2014; Priem, Piwowar, et al., 2012). Lower rates have been found for publications from particular geographic regions, such as Iran (Maleki, 2014), Brazil, and Latin-American countries (Alperin, 2015), while higher rates have been found for certain journals (Eysenbach, 2011) and groups of papers—such as those submitted to arXiv (Haustein, Bowman, Macaluso, et al., 2014; Shuai, Pepe, & Bollen, 2012)—which can be partly explained by automated bot accounts (Haustein, Bowman, Holmberg, et al., 2016). Large differences are also found on the coverage of particular datasets over time: for example, 12% of PLOS papers were tweeted at least once as of 2012 (Priem, Piwowar, et al., 2012), this percentage increased to 53% in 2015 (Barthel et al., 2015).

Correlational studies between Twitter mentions and citations have found mixed results: large-scale studies have revealed correlations between .1-.2 (Barthel et al., 2015; Costas et al., 2014; Haustein, Peters, Sugimoto, et al., 2014; Priem, Piwowar, et al., 2012), while discipline or journal samples have shown higher correlations (Eysenbach, 2011; Shuai et al., 2012). Number of tweets per paper has been shown to be at 0.78 (Haustein, Costas, et al., 2015) for articles in WoS and 2.82 for PLOS papers (Barthel et al., 2015), with intensity (i.e., excluding non-tweeted papers) between 2.5 (Haustein, Peters, Sugimoto, et al., 2014) and 3.65 (Haustein, Costas, et al., 2015). Twitter activity differs largely between disciplines with papers from the social sciences and biomedical and health sciences receiving more twitter attention than those form mathematics and computer science, and natural sciences and engineering (Costas et al., 2014, 2015; Haustein, Costas, et al., 2015).

As one might expect from this medium, the "twitter window" of scientific papers is quite short, with an increase following publication and fast decay (Eysenbach, 2011; Shuai et al., 2012). This has led scholars to suggest that tweets can serve as early predictors of citations. However, some evidence complicates this narrative: in a study of papers published in 2012, only one-fifth had been tweeted, whereas two-thirds of the articles had been cited by the end of 2013 (Haustein, Costas, et al., 2015). Therefore, while most papers are cited following a certain number of years, most papers remain untweeted.

Studies have also sought to examine what type of content receives higher number of tweets: for example, meta-analyses, systematic reviews, and clinical trials were found to be tweeted more frequently than other study types (Andersen & Haustein, 2015). There is also a relationship between article's length and their number of tweets, with shorter articles, editorials, and news items receiving more tweets than longer ones (Haustein, Costas, et al., 2015). Productivity studies have examined the relationship between researchers' output in terms of microblogging and other forms of research output (e.g., writing journal articles). A negative correlation has been found, suggesting that those who tweet tend to publish less (Haustein, Bowman, Holmberg, et al., 2014). Such an assumption led to the development of a tongue-in-cheek index—the Kardashian Index (K-index)—which was developed to the determine the magnitude of a scientists' Twitter popularity in relation to their scientific output (Hall, 2014). Other studies have found no relationship between the number of publications and number of followers (Mas-Bleda et al., 2014).

**Wikis**

Although most scholars do not create content on Wikipedia (Procter et al., 2010b), it is increasingly cited, both in journal articles (Rousidis et al., 2013) and in scholarly blog posts (Weller & Peters, 2012). The percentage of articles cited by Wikipedia pages remains low: the few studies of coverage of articles in Wikipedia have shown rates of coverage around 1% (Alperin, 2014, 2015; Zahedi, Costas, & Wouters, 2014b), though the rate of coverage is slightly higher in the case of computer science (3%) (Shuai, Jiang, Liu, & Bollen, 2013) and PLOS papers (~5%) (Fenner, 2014; Lin & Fenner, 2014; Priem, Piwowar, et al., 2012).

An early analysis of Wikipedia citations found correlations between citations and mentions on Wikipedia around .25 (F. Å. Nielsen, 2007). As with other social media metrics, high-impact journals—e.g., *Nature*, *Science*, and the *NEJM*—were overrepresented (F. Å. Nielsen, 2007). Subsequent studies have also examined the citation advantage for articles on Wikipedia, finding that articles referenced in Wikipedia tended to have higher citation counts than a control sample (Evans & Krauthammer, 2011). However, this was described as a selectivity bias by a subsequent study, which did not show an increase in propensity to be cited, based on citation in Wikipedia (Marashi et al., 2013). Such studies reinforce the perceptions of scholars: two-thirds of them did not believe that mentions of their work in Wikipedia could be used for evaluation purposes (Haustein, Peters, Bar-Ilan, et al., 2014). Positive relationships have been found between Wikipedia impact and other metrics of scholarly success (Shuai et al., 2013; Chen, Tang, Wang, & Hsiang, 2015) have been demonstrated for some populations. However, others have suggested that high research performance does not explain presence on Wikipedia (Samoilenko & Yasseri, 2014).

**Social recommending, rating and reviewing**

One of the characteristics of social media that has been heralded is the opportunity to create dynamic and crowdsourced evaluations of research in ways that are faster and more equitable than traditional peer review. Examples of such social platforms are Faculty of 1000 (F1000Prime) and thirdreviewer.com (Mandavilli, 2011). Post-peer review sites like these ones have been likened to conversation at a research conference (Faulkes, 2014) and the evaluative aspects have been rated above their review potential (Wets, Weedon, & Velterop, 2003). On F1000Prime, papers are rated as "Good", "Very Good", or "Exceptional" and assigned a numeric rating based on the number of reviews and ratings assigned (FFa score) and can be rated multiple times, though most papers only have a single rating (Waltman & Costas, 2014). Reviewers may also leave comments explaining their rating and tag a publication with labels such as "Controversial", "Technical Advance", "Refutation", etc.

Studies of the coverage of F1000 are relatively rare, with rates ranging from less than 1% (Knight, 2014) to 2-8% (Bornmann & Leydesdorff, 2013; Lin & Fenner, 2014; Priem, Piwowar, et al., 2012). Correlations between citation counts and F1000 hover around .3-.4 (Bornmann & Leydesdorff, 2013; Eyre-Walker & Stoletzki, 2013; Li & Thelwall, 2012; Mohammadi & Thelwall, 2013; Waltman & Costas, 2014), with variations by number of ratings (Waltman & Costas, 2014), Impact Factor (Anonymous, 2005), and tag type (e.g., "New Findings", "Changes to clinical practice") (Mohammadi & Thelwall, 2013). Tags were also influential in predicting shares on other social media sites (Bornmann, 2014b, 2015). Correlations between expert panel ratings and F1000 have shown to be moderately positive (.445); however, a number of papers identified as major contributions were not reviewed by F1000, showing that the site may not be comprehensive in identifying influential papers (Allen, Jones, Dolby, Lynn, & Walport, 2009; Wardle, 2010). There is also a relationship with articles mentioned on Wikipedia and found in F1000 (Evans & Krauthammer, 2011).

Other platforms with reviews and ratings have also been exploited as potential research evaluation tools, particularly for monograph-based disciplines. In a study of monographs indexed in the Thomson Reuters Book Citation Index, 29% of the books were found to have reviews on Amazon.com (Kousha & Thelwall, 2015). Arts and Humanities had proportionally more books with reviews (Kousha & Thelwall, 2015), suggesting this might be a good source for disciplines historically underrepresented in bibliometric databases (Cronin & Sugimoto, 2015). However, correlations between Amazon reviews/stars and citation counts was quite low, between .1-.2, depending on the discipline (Kousha & Thelwall, 2015). Similar correlations were found in a study of Goodread ratings of history books (Zuccala et al., 2015). Reddit has been identified as a potential source, though low rates of coverage (Htoo & Na, 2015) and inconclusive correlation results (Htoo & Na, 2015; Thelwall, Haustein, et al., 2013) have been found so far.

# 4 Conclusion and outlook

The advent of the digital era and, more specifically, of social media, has yielded the emergence of new online tools that allow for diffusing, discussing, and organizing scholarship, as well as a new family of research indicators to measure these activities. This review aimed at providing a comprehensive summary of the empirical literature on these practices and indicators, with an emphasis on the roles of the various platforms for scholars and their organizations, and on the strengths and limitations of altmetric indicators. A central theme of this review is heterogeneity: not only do the results obtained often diverge from one study to another due to different methods, samples, or differences between the time of analysis, but the acts on the social media platforms underlying various altmetrics are extremely heterogeneous. Therefore, findings regarding altmetrics as well as the use of social media in academia are difficult to generalize. We have thus tried to summarize the current literature by identifying several types of acts performed on various social media platforms: social networking, social bookmarking, social data sharing, video, blogging, microblogging, as well as social recommending, rating and reviewing. However, due to the ever-changing landscape of social media platforms, which through particular affordances generate new online acts and traces, a similar review performed in a few years might provide drastically different results. This serves, therefore, as a state-of-the-art.

The number of papers reviewed provides evidence of the popularity and interest social media and associated indicators have generated in the scientific community. The technological push, initiated by start-ups and scholarly publishers, is met by a policy pull of funders and researchers demanding indicators that reflect academic productivity and influence beyond papers published and citations received. The fast emergence and adoption of altmetrics can therefore be interpreted as a technologically enabled convergence of the interests of these stakeholders, which took place on the fertile ground of the current evaluation culture of academe—and of contemporary society in general (Dahler-Larsen, 2011). Indeed, this credit-driven academic culture demands a reward for any type of contribution made—necessitating and incentivizing the tracking and recording the entire spectrum of scholarly acts. Rewarding social media activities creates incentives for scholars to use these platforms, potentially leading to *gamification* of research activities (following Campbell's law) and large-scale goal displacement.

The dependency of altmetric indicators on underlying social media platforms cannot be emphasized enough. These platforms and their affordances allow for the online acts to take place and influence the way in which they are performed. In fact, most of these acts cannot exist outside of the particular platform, which translates into a variety of indicators *entirely* specific to and dependent on the underlying tool. It has argued that social media "platforms" sociality (Alaimo & Kallinikos, In-press)—by extension, it could be argued that the rise of altmetrics allows social media to also begin to platform science. This dependency explains, at least to some extent, the difficulty of defining the concepts behind altmetrics. Similar arguments have been made about citation indexes and the transfiguration of the scientific object (e.g., the reference or citation) when it is platformed in particular ways (Day, 2014; Wouters, 2016). In this context, stakeholders that supply and demand altmetrics and the use of social media in academia must be cautious not to take the shadow for the substance, where measurable traces of research activities and impact become more important than the activities themselves.

Despite these blind spots and potential curves in the road ahead, we would argue that the increased use of social media in scholarly communication and the adoption of altmetrics are more than simple fads. While the first wave of digitization of scholarly communication—in which we include emails, listservs, as well as electronic journals—translated into faster discussions *within* the scientific community, this second wave of digitization includes the use of tools that do allow for broader discussion *outside* the scientific community and, thus, could allow for a broader conversation about research. But the presence of these platforms alone does not guarantee broader conversation: initial studies suggest that social media has rather opened a new channel for informal discussions among researchers, rather than a bridge between the research community and society at large. Leveraging the power of social media to achieve the latter will require a careful negotiation between vendors, institutions, and policy makers.

Time will tell whether social media and altmetrics are an epiphenomenon of the research landscape, or if they become central to scholars' research dissemination and evaluation practices. While some indicators and platforms might disappear because they lack meaning or relevance, others might share have the same fate due to the termination of a platform or service on which they are based, or for which they were the sole source which, again, highlights the strong—if not entire—dependency of these indicators and practices on their platforms.

# Acknowledgements

This review was supported by the Alfred P. Sloan Foundation Grant #G-2014-3-25, by the Social Sciences and Humanities Research Council of Canada, as well as from the Canada Research Chairs program.

# Cited References

submitted to JASIST